\documentclass[11pt,a4paper]{article}
\usepackage{amsmath,amssymb,amsthm,amsfonts,mathtools}
\usepackage{physics}
\usepackage[dvips]{graphicx}
\usepackage{subfigure}
\usepackage{verbatim}
\usepackage[font=small]{caption}
\usepackage{soul}
\usepackage{commath}
\usepackage{hyperref}
\hypersetup{colorlinks={true},linkcolor={blue},urlcolor=blue,citecolor=green,linktocpage=true}
\usepackage{color}
\makeatletter
\let\@fnsymbol\@arabic
\makeatother

\usepackage{rotating}

\title{Magnetic nanodrug delivery in non-Newtonian blood flows}

\author{C. Fanelli\thanks{Departament de Matem\`{a}tica Aplicada, Universitat Polit\`{e}cnica de Catalunya - BarcelonaTech, Barcelona, Spain.}, K. Kaouri\thanks{School of Mathematics, Cardiff University, CF24 4AG, UK}, T.N. Phillips\footnotemark[2], T.G. Myers\footnotemark[1] $^,$\thanks{Centre de Recerca Matem\`{a}tica, Campus de Bellaterra  Edifici C, 08193 Bellaterra, Barcelona, Spain.}, F. Font\thanks{Department of Fluid Mechanics, Universitat Polit\`{e}cnica de Catalunya - BarcelonaTech, Barcelona, Spain.} 
$^,$\footnote{francesc.font@upc.edu}}

\begin{document}
\maketitle

\abstract{With the goal of determining strategies to maximise drug delivery to a specific site in the body, we developed a mathematical model for the transport of drug nanocarriers (nanoparticles) in the bloodstream under the influence of an external magnetic field. Under the assumption of long (compared to the radius) blood vessels the Navier-Stokes equations are reduced, to a simpler model consistently with lubrication theory.  Under these assumptions, analytical results are compared for Newtonian, power-law, Carreau and Ellis fluids, and these clearly demonstrate the importance of shear thinning effects when modelling blood flow. Incorporating nanoparticles and a magnetic field to the model we develop a numerical scheme and study the particle motion for different field strengths. We demonstrate the importance of the non-Newtonian behaviour: for the flow regimes investigated in this work, consistent with those in blood micro vessels, we find that the field strength needed to absorb a certain amount of particles in a non-Newtonian fluid has to be larger than the one needed in a Newtonian fluid. Specifically, for one case examined, a two times larger magnetic force had to be applied in the Ellis fluid than in the Newtonian fluid for the same number of particles to be absorbed through the vessel wall. Consequently, models based on a Newtonian fluid can drastically overestimate the effect of a magnetic field. Finally, we evaluate the particle concentration at the vessel wall and compute the evolution of the particle flux through the wall for different permeability values, as that is important when assessing the efficacy of drug delivery applications. The insights from our work bring us a step closer to successfully transferring magnetic nanoparticle drug delivery to the clinic.}


\section{Introduction}
\label{sect:introduction}

Currently, the main approaches in cancer therapy include surgery, chemotherapy, radiotherapy, and hormone therapy \cite{bahrami2017nanoparticles}. The first is invasive while the latter three are non-specific. Hence, their efficacy is not only low but sometimes the cure can be more aggressive than the disease itself. In the late 1970s, magnetic drug targeting, which consists of injecting and steering magnetic drug carriers through the vessel system towards disease locations using an external magnetic field, was proposed as an alternative and more efficient treatment for tumors (see \cite{pankhurst2003applications} and references therein).  At present, there are many promising studies, both \textit{in vivo} and \textit{in vitro} \cite{alexiou2005invitro, muthana2008novel,wang2017theranostic} but, to our knowledge, only a few successful trials on human patients have been carried out \cite{lubbe1996clinical,wilson2004hepatocellular,lemke2004mri}. All studies demonstrate that magnetic forces can attract particles in the region near the magnet but there is a lack of knowledge on how to quantify and optimise the accumulation of particles \cite{fiocchi2019computational}.

Describing the movement of particles subject to a magnetic field in the bloodstream is a relatively difficult task due to the interplay of magnetic and hydrodynamic forces acting on the particles and  the inherent difficulty of solving the Navier-Stokes equations. One further difficulty is to accurately model and simulate the non-Newtonian behaviour of blood. Experiments show that for low shear rates ($\sim$1\,s$^{-1}$), the viscosity can be as high as 10-11\,mPa$\cdot$s while for shear rates in excess of $1000$\,s$^{-1}$, it tends to an asymptotic value of 3-4\,mPa$\cdot$s \cite{Yam2020, fournier2017basic}. In order to simplify calculations, previous studies consider blood as a Newtonian fluid ~\cite{grief2005mathematical,richardson2010particle,yue2012motion,boghi2017numerical} or as a non-Newtonian power-law fluid \cite{nacev2011behaviors,cherry2014comprehensive,lunoo2015capture,rukshin2017modeling}. Grief and Richardson \cite{grief2005mathematical} and then Richardson \textit{et al.} \cite{richardson2010particle} developed a continuum model for the motion of particles subject to a magnetic field, both using a Newtonian flow model for blood. In \cite{grief2005mathematical} it was shown, via a simple network model, that it is impossible to specifically target interior regions of the body with an external magnetic field; the magnet can be used only for targets close to the surface. In \cite{richardson2010particle} the boundary layer structure in which particles concentrate was analysed. Using a similar approach, Nacev \textit{et al.} \cite{nacev2011behaviors} simulated particle behaviour under the influence of a magnetic field using a power-law assumption for the blood flow. Cherry and Eaton \cite{cherry2014comprehensive} developed a comprehensive continuum model for the motion of micro-scale particles using a value for the viscosity determined from fitting experimental data \cite{brooks1970interactions} to model blood shear thinning, and used the model to investigate magnetic particle steering through a branching vessel. 

A structural difference within the models developed in the literature lies in how particles are described. Instead of describing particle distribution in blood as a continuum and using a diffusion equation to study the evolution of the particle concentration, several studies consider particles as discrete elements and track the trajectory of each individual particle \cite{yue2012motion,boghi2017numerical,lunoo2015capture,rukshin2017modeling,Freund2012computational,ye2018computational}. Yue \textit{et al.} \cite{yue2012motion} implemented a stochastic model for the transport of nanoparticle clusters in a Hagen-Poiseuille flow in order to find an optimal injection point. Using a similar approach, Rukshin \textit{et al.} \cite{rukshin2017modeling} developed a stochastic model to simulate the behaviour of magnetic particles in small vessels. Lunnoo and Puangmali \cite{lunoo2015capture} used a generalized power-law model to investigate the parameters which play a crucial role in magnetic drug targeting, showing how difficult it can be to keep small particles in the desired region. Recently, in Boghi \textit{et al.} \cite{boghi2017numerical} a numerical simulation of drug delivery in the blood in the coeliac trunk was performed.

In the present work we analyse the forces and parameters involved in the process of magnetic drug delivery and highlight the importance of considering realistic non-Newtonian models for blood flow and the motion of the nanoparticles in it. It is crucial to consider the non-Newtonian behaviour of the blood in order to predict if particles are able to reach the desired area. A mathematical model in a two-dimensional channel is introduced in Section \ref{sect:genericmathmodel}. In Section \ref{sec:bloodnonnewt}, we explain how choosing an oversimplified constitutive law for the fluid in the vessel or an incorrect value for the viscosity in the centre of the channel can lead to significant errors. In Section \ref{sec:conceq} we demonstrate how magnetic forces act on particles depending on their size. Numerical simulations illustrating the influence of key parameters are presented in Section \ref{sec:results}. We draw our conclusions in Section \ref{sec:conclu}.

\section{Governing equations for non-Newtonian blood flow}
\label{sect:genericmathmodel}
\begin{figure}[h]
	\centering
	\includegraphics[width=\linewidth]{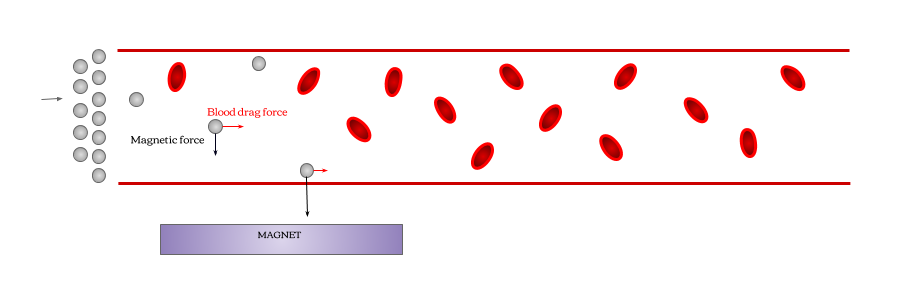}
	\caption{Sketch of the injection of magnetic nanoparticles in a vessel, in the presence of red blood cells and subject to a magnetic field.}
	\label{fig:vessel_Scheme}
\end{figure}

As illustrated in Figure \ref{fig:vessel_Scheme}, we approximate the vessel as a long and thin rectangular channel, consistent with the size of vessels in the human body (see Table~\ref{Table:sizevessels}). The particles are injected at the inlet of the vessel and their motion is driven by the  field generated by an external magnet located at the bottom of the domain and by the drag force on the particles due to the blood flow.

\begin{table}[h]
	\centering
	\begin{tabular}{lccc}
		\hline
		\noalign{\vskip 2mm}
		\textbf{Vessel} & $L$ (m)  & $R$ (m) & $N$\\
		\noalign{\vskip 2mm}
		\hline
		\noalign{\vskip 2mm}
		Aorta & $0.4$ & $1.25 \times 10^{-2}$ & 1\\
		Artery & $0.1$ & $1.5 \times 10^{-3}$ & 159 \\
		Arteriole & $7 \times 10^{-4}$ & $2.5 \times 10^{-5}$ & $5.7 \times 10^7$ \\
		Capillary & $6 \times 10^{-4}$ & $4 \times 10^{-6}$ & $1.6 \times 10^{10}$ \\
		Venule & $8 \times 10^{-4}$ & $1 \times 10^{-5}$ & $1.3 \times 10^9$\\
		Vein & $0.1$ & $2.5 \times 10^{-3}$ & 200\\
		Vena cava & $0.22$ & $1.5 \times 10^{-2}$ & 2\\
		\noalign{\vskip 2mm}
		\hline
	\end{tabular}
	\caption{Typical values for the various types of vessels in human body: average length ($L$) and radius ($R$), estimation of the average number of a specific vessel in the circulatory system ($N$). All values are adapted from \cite{formaggia2009cardiovascular}.}
	\label{Table:sizevessels}
\end{table}

Blood containing nanoparticles can be considered as a nanofluid. Typically, its motion is governed by the Navier-Stokes equations coupled to an advection-diffusion equation for the concentration of particles in the fluid \cite{myers2017NDebate}. The Navier-Stokes equations describing the flow of an incompressible nanofluid in a vessel subject to an external magnetic field are
\begin{align}
\rho \left[ \frac{\partial \textbf{u}_F}{\partial  t} + \textbf{u}_F \cdot \nabla \textbf{u}_F \right] &  = - \nabla p + \nabla \cdot \boldsymbol\tau + \textbf{F}_{\text{mag}}, \label{eqn:momentum_magn_cart} \\
\nabla \cdot \textbf{u}_F & = 0, \label{eqn:massconservation_magn_cart}
\end{align}
where $\mathbf{u}_F$ is the fluid velocity, $\rho$ is the fluid density, $p$ is pressure, $\boldsymbol\tau$ is the extra-stress tensor and $\textbf{F}_{\text{mag}}$ the magnetic force acting on the fluid.

In this work, we consider the nanofluid, composed of blood and nanoparticles, to be a dilute mixture. This allows us to make two assumptions which substantially reduce the complexity of \eqref{eqn:momentum_magn_cart}-\eqref{eqn:massconservation_magn_cart}. Firstly, the density of the nanofluid, $\rho$, which typically depends strongly on the particle concentration \cite{macdevette2014boundary,myers2017NDebate}, can be considered approximately constant and equal to the density of blood. Secondly, given that blood has negligible magnetization \cite{alimohamadi2014transient} and its overall character is found to be paramagnetic \cite{voltairas2002hydrodynamics}, the magnetic field only acts on the nanoparticles. Since we are considering a dilute mixture, the capacity of nanoparticles to modify the flow rheology is negligible and therefore we can assume $\boldsymbol{F}_{\text{mag}} \approx 0$ in \eqref{eqn:momentum_magn_cart}-\eqref{eqn:massconservation_magn_cart}. Due to these assumptions, the flow equations \eqref{eqn:momentum_magn_cart}-\eqref{eqn:massconservation_magn_cart} are effectively decoupled from the mass conservation (advection-diffusion) equation that will describe the concentration of particles within the vessel (see Section~\ref{sec:conceq}).

\subsection{Blood as a non-Newtonian fluid}
\label{sec:bloodnonnewt}
Bodily fluids, such as blood, saliva, and eye fluid are invariably non-Newtonian. In particular, blood is a concentrated suspension of particles in plasma, which is mainly made of water. The three most important 'particles' that constitute blood are: red blood cells (RBCs), white cells and platelets. RBCs, which are the most numerous, are  mainly responsible for the mechanical properties of blood  \cite{formaggia2009cardiovascular}. In fact, their tendency to form (and then break down) three-dimensional microstructures at low shear rates and to align to the flow at high shear rates cause the blood's shear thinning behaviour, characterized by the monotonic decrease of the viscosity that tends to some limit for very high shear rates \cite{fournier2017basic}. In the case of blood, the  structures lead to significant changes in its rheological properties and several models have been developed during the past 50 years in order to capture the complexity of this behaviour (some examples can be found in \cite{ballyk1994simulation, chien1970shear, cho1991effets, huang1976thixotropic, kaliviotis2018local, sherwood2014spatial}). However, none of those models has been universally accepted.

In mathematical terms, we define a fluid as non-Newtonian if the extra-stress tensor cannot be expressed as a linear function of the components of the velocity gradient. The more general relation between the stress and rate-of-strain tensors can be written as
\begin{equation}\label{eqn:shear}
\boldsymbol\tau = \eta (\dot{\gamma}) \dot{\boldsymbol{\gamma}},
\end{equation}
where $\eta (\dot{\gamma})$ is the viscosity, $\dot{\boldsymbol{\gamma}}=\nabla \mathbf{u} + \nabla \mathbf{u}^T$ and
\begin{equation} \label{eqn:shearrate}
\dot{\gamma} = \left[ \frac{1}{2} \left( \frac{\partial u_i}{\partial  x_j} + \frac{\partial u_j}{\partial  x_i} \right)\left( \frac{\partial u_i}{\partial  x_j} + \frac{\partial u_j}{\partial  x_i} \right)\right]^{1/2}\,,
\end{equation}
is the generalised shear rate. When
$\eta (\dot{\gamma})$ is constant the Newtonian model is recovered. A purely shear-thinning fluid will exhibit a monotonic decrease in viscosity with increasing shear rate. Practical fluids are more likely to exhibit a constant viscosity beyond certain high and low values of shear rate called \lq Newtonian plateaus'. In between, a nonlinear viscosity relation links the two plateaus. 

The aim of this section is to compare different types of simple, non-Newtonian fluid models. In particular, we will compare the Newtonian model with the power-law model, the Carreau model and the Ellis model. In Tables~\ref{nonnewmodels} and Table~\ref{Table:PhysicalParameters_flow} we summarize the expressions for the viscosity and shear stress for each model and the corresponding parameter values, respectively. Note that under the assumption of flow in a long thin channel, we will significantly simplify the governing equations (Section \ref{simpli_flow_eqs}). The shear stress relations in Table~\ref{nonnewmodels} are consistent with this reduction. In the power-law model $m$ is constant. If $n_p<1$ the fluid is pseudoplastic or shear thinning and if $n_p>1$ it is dilatant or shear thickening; for blood $n_p=0.357<1$. If $n_p=1$ we retrieve the Newtonian expression. In the Carreau model $\lambda$ is a constant and $\eta_0$ and $\eta_\infty$ are the limiting viscosities at low and high shear rates, respectively. In the Ellis model $\eta_0$ is the viscosity at zero shear and $\tau_{1/2}$ is the shear stress at which the viscosity is $\eta_0/2$. The latter model does not include a high viscosity plateau, however for most practical situations such high strain rates are never reached and so this plateau value is not relevant. In the case of standard blood flow we do not anticipate high shear rates.

\begin{table}[h]
	\centering
	\begin{sideways}
	\begin{tabular}{lll}
		\hline
		\textbf{Model} & \textbf{Viscosity} & \textbf{Shear stress} \\
		\hline \\
		Newtonian   & $\mu=$ constant
		& $\tau_{yx}  =  \mu \abs{\frac{\partial u}{\partial y}}$ \\
		\noalign{\vskip 2mm} 		
		\hline \\
		Power-law  & $\eta_p(\dot{\gamma})= m\,\abs{\dot{\gamma}}^{n_p -1}$ & $\tau_{yx}=  m\,\abs{\frac{\partial u}{\partial y}}^{n_p}$ \\
		\noalign{\vskip 2mm}
		\hline \\
		Carreau    & $	\eta_c (\dot{\gamma}) = \eta_\infty + (\eta_0 - \eta_\infty) \left(1 + \lambda^2 \dot{\gamma}^2 \right)^{\frac{n_c -1}{2}} $
		& $\tau_{yx}  = \eta_\infty + \left(\eta_0 - \eta_\infty \right)  \left(1 + \lambda^2 \abs{\frac{\partial u}{\partial y}}^2 \right) ^{\frac{n_c-1}{2}} \abs{\frac{\partial u}{\partial y}}$ \\
		\noalign{\vskip 2mm}
		\hline \\
		Ellis      & $	\eta_e (\tau) = \eta_0 \left(1 + \abs{\frac{\tau}{\tau_{1/2}}}^{\alpha - 1} \right)^{-1}$
		& $\eta_e(\tau_{yx})= \eta_0 \left(1 + \abs{ \frac{\tau_{yx}}{\beta}}^{\alpha - 1} \right)^{-1}$  \\
		\noalign{\vskip 2mm} 		
		\hline
	\end{tabular}
	\end{sideways}
	\caption{Newtonian and non-Newtonian viscosity models and corresponding shear stress, assuming $\dot{\gamma}\approx \abs{\partial u/\partial y}$.}
	\label{nonnewmodels}
\end{table}

\begin{table}[h]
	\centering
	\begin{tabular}{lcccc}
		\hline
		\textbf{Quantity} & \textbf{Symbol} & \textbf{Value} & \textbf{Units} & \textbf{References} \\
		\hline
		\noalign{\vskip 2mm}
		Newtonian viscosity & $\mu$ & $0.0035$ & Pa s  & \cite{johnston2004nonnewt} \\
		\noalign{\vskip 2mm}
		\noalign{\vskip 2mm}
		Power-law viscosity   & $m$ & 0.035 & Pa s& \cite{myers2005application} \\
		Power-law exponent & $n_p$ & 0.357 & No. & \cite{myers2005application} \\
		\noalign{\vskip 2mm}
		\noalign{\vskip 2mm}
		Carreau coefficient & $\lambda$ & 3.313 & No. & \cite{fournier2017basic} \\
		Carreau viscosity at low shear rates & $\eta_0$ & 0.056 & Pa s   & \cite{fournier2017basic} \\
		Carreau viscosity at high shear rates & $\eta_\infty$ & 0.0035 & Pa s & \cite{fournier2017basic} \\
		Carreau exponent & $n_c$ & 0.357 & No. & \cite{fournier2017basic}  \\
		\noalign{\vskip 2mm}
		\noalign{\vskip 2mm}
		Ellis viscosity at low shear rate & $\eta_0$ & $0.056$ & Pa s  & \cite{myers2005application}\\
		Ellis shear stress at $\eta_0/2$ & $\beta$ & 0.026 & Pa  & \cite{myers2005application} \\
		Ellis exponent & $\alpha$ & 3.4 & No.  & \cite{myers2005application} \\
		\noalign{\vskip 2mm}
		\hline
	\end{tabular}
	\caption{Typical parameter values for the blood flow equations for each model, taken from \cite{johnston2004nonnewt,myers2005application,fournier2017basic}.}
	\label{Table:PhysicalParameters_flow}
\end{table}

\begin{figure}[h]
	\centering
	\includegraphics[width=0.6\linewidth]{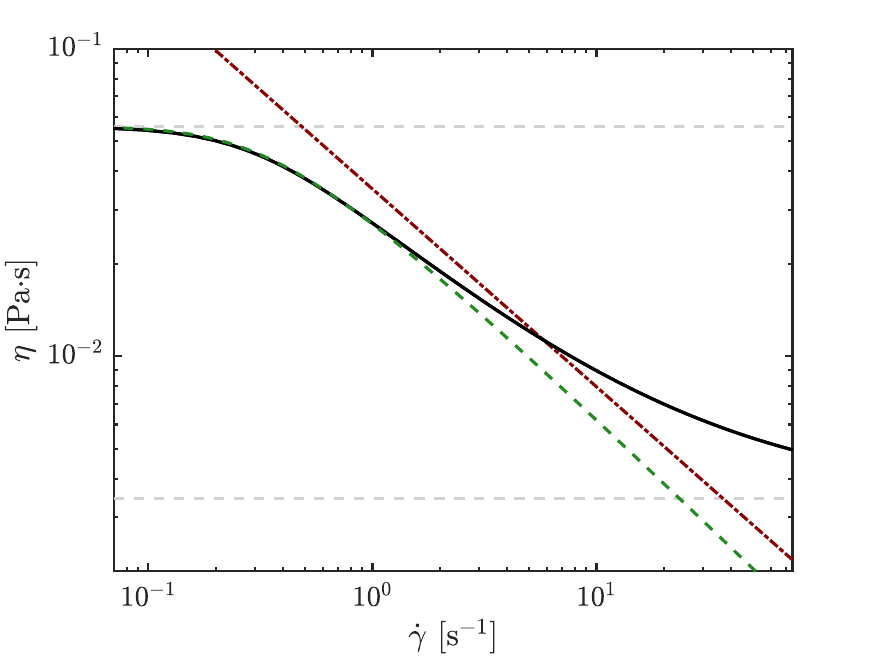}
	\caption{The viscosity/shear rate plot in logarithmic scale for the power-law (red dotted-dashed line), the Carreau (black solid line) and the Ellis models (green dashed line). The light grey dashed lines represent the limiting viscosities $\eta_0$ and $\eta_\infty$.}
	\label{fig:non-newtonian-behaviour}
\end{figure}

In Figure \ref{fig:non-newtonian-behaviour} we compare the behaviour of the four models, using the parameter values in Table~\ref{Table:PhysicalParameters_flow}. For moderate shear rates it is clear that the Ellis and Carreau models show good agreement but differences emerge as the shear rate increases above $\dot{\gamma} \approx 1.6$ $s^{-1}$. In attempting to compensate for the plateau values the power-law model seems to be inaccurate over the whole range of shear rates. 

In the next section using the above models we determine the velocity profiles and demonstrate the importance of choosing the right model for blood transporting nanoparticles under the influence of an external magnetic field.

\subsection{Nondimensionalisation and simplification of flow equations}\label{simpli_flow_eqs}

Expressing equations \eqref{eqn:momentum_magn_cart}--\eqref{eqn:massconservation_magn_cart} in components, the governing equations for the fluid become
\begin{align}
\frac{\partial u}{\partial  x} + \frac{\partial v }{\partial y} &  = 0, \label{eqn:massconservation_cart} \\
\frac{\partial u}{\partial  t} +u \frac{\partial u }{\partial x } + v \frac{\partial u}{\partial  y} &  = -\frac{1}{\rho}\frac{\partial p}{\partial  x} +\frac{1}{\rho} \left[\frac{\partial \tau_{xx}}{\partial  x} +\frac{\partial  \tau_{yx}}{\partial y} \right], \label{eqn:momentum1_cart} \\
\frac{\partial v}{\partial  t} +u \frac{\partial v }{\partial x } +v  \frac{\partial v}{\partial  y} & = -\frac{1}{\rho}\frac{\partial p}{\partial y} + \frac{1}{\rho } \left[ \frac{\partial  \tau_{xy}}{\partial  x} +\frac{\partial  \tau_{yy}}{\partial y} \right]. \label{eqn:momentum2_cart}
\end{align}
which are subject to the no-slip conditions 
\begin{align}
u = 0\,,\quad v = 0\quad \text{at}\ & y = \pm R\,, \label{bcnoslip}
\end{align}
and to the symmetry condition
\begin{align}
\frac{\partial u}{\partial y}=0\quad \text{at}\ & y=0\,. \label{bcsymm}
\end{align}

In order to determine the order of magnitude of the terms in equations (\ref{eqn:massconservation_cart})--(\ref{eqn:momentum2_cart}), we proceed to non-dimensionalise. We define the non-dimensional variables
\begin{equation}\label{eqn:nondimvariab_cart}
x=L \hat{x}, \quad y= 2R \hat{y}, \quad u=U \hat{u}, \quad v = V \hat{v}, \quad p =P \hat{p}, \quad \tau = \mathcal{T} \hat{\tau} \,.
\end{equation}

In (\ref{eqn:massconservation_cart}) we can balance the terms by setting $V = \varepsilon U$, where $\varepsilon = 2R/L \ll 1$ (since the blood vessels are long and thin). Hence, the non-dimensional continuity equation becomes
\begin{equation}\label{eqn:balancecart_nd}
\frac{\partial \hat{u}}{\partial  \hat{x}} +\frac{\partial \hat{v}}{\partial  \hat{y}}   = 0.
\end{equation}
Choosing $P = L \mathcal{T} /2R$ and assuming steady-state, we can write
\begin{align}
\frac{\rho \varepsilon U^2 }{\mathcal{T}} \left[ \hat{u} \frac{\partial \hat{u}}{\partial  \hat{x}}  + \hat{v} \frac{\partial \hat{u} }{\partial \hat{y} } \right] &= - \frac{\partial \hat{p}}{\partial \hat{x}}  + \varepsilon \frac{\partial  \hat{\tau}_{\hat{x}\hat{x}} }{\partial \hat{x}} + \frac{\partial  \hat{\tau}_{\hat{y}\hat{x}} }{\partial \hat{y}}\,, \label{NS_non1}\\
\frac{\rho \varepsilon U^2 }{\mathcal{T}} \left[ \hat{u}\frac{\partial \hat{v}  }{\partial \hat{x} } +\hat{v}   \frac{\partial \hat{v} }{\partial \hat{y} } \right] &= - \frac{\partial \hat{p}}{\partial \hat{y}} + \varepsilon^2 \frac{\partial  \hat{\tau}_{\hat{x}\hat{y}}}{\partial  \hat{x}} + \varepsilon \frac{\partial  \hat{\tau}_{\hat{y}\hat{y}}}{\partial \hat{y}}\,. \label{NS_non2}
\end{align}
Assuming $\frac{ \rho \varepsilon U^2}{\mathcal{T}} \ll 1$  and neglecting terms of $O(\varepsilon)$ or smaller in (\ref{NS_non1})--(\ref{NS_non2}),  the resulting governing equations, in dimensional form, are
\begin{align}
& \frac{\partial u}{\partial  x} + \frac{\partial v}{\partial  y}= 0\,, \label{eqn:cont_lub_cart} \\
&  \frac{\partial p}{\partial x} = \frac{\partial  \tau_{yx} }{\partial y}\,, \label{eqn:momentum1_lub_cart} \\
& \frac{\partial p}{\partial y}  = 0\,, \label{eqn:momentum2_lub_cart}
\end{align}
which is a much simpler system of equations, consistent with lubrication theory \cite{ockendon_ockendon_1995}. 

\subsection{Comparison of four viscosity models}
\label{subsec:comparison}

To determine the most accurate model describing blood flow, we now compare the fluid velocity profiles obtained for the four viscosity laws summarised in Table~\ref{nonnewmodels}. In all cases we consider a Hagen-Poiseuille flow in a channel, which is driven by a constant pressure gradient, $\Delta p/L$, and has a constant volumetric flow rate, $Q$. The results quoted below follow from the models in \cite{myers2005application} for flows driven by a pressure gradient and a moving bottom surface, where we set the velocity of the surface to zero. To switch to the current vertical coordinate $y$ from that used in \cite{myers2005application} we set $z = (y+R)h/(2R)$. The position of the turning point in the flow,  $z=z_m=h/2$, then corresponds to $y=0$.

\paragraph{The Newtonian  model:} The solution for the Newtonian fluid is obtained by solving equations (\ref{eqn:cont_lub_cart})--(\ref{eqn:momentum2_lub_cart}) where the shear stress is specified in Table~\ref{nonnewmodels}. From equation (\ref{eqn:momentum2_lub_cart}), the pressure does not vary with $y$. Hence, integrating (\ref{eqn:momentum1_lub_cart}) twice with respect to $y$ and using the boundary conditions \eqref{bcnoslip}-\eqref{bcsymm} we obtain the well-known parabolic profile
\begin{equation}\label{eqn:velocitynewtonian}
u(y) =\frac{3Q}{4 R^3}\left(R^2 - y^2\right)\,,
\end{equation}
where the relation between $\partial p/\partial x$ and $Q$ is determined using that $Q=\int_{-R}^{R}\,u\,dy=\text{constant}$  \cite{myers2005application}.

\paragraph{The power-law model:} Proceeding similarly to the Newtonian case, the  velocity for a non-Newtonian power-law fluid with the corresponding shear stress from Table~\ref{nonnewmodels} becomes
\begin{equation}
u(y) = \frac{Q}{R^{\frac{2n_p+1}{n_p}}} \left( \frac{2n_p+1}{2n_p+2} \right) \left( R^{\frac{n_p+1}{n_p}} -\abs{y}^{\frac{n_p+1}{n_p}}\right).
\end{equation}
The main disadvantage of this  model is that the viscosity $\eta \propto 1/(\partial u/\partial y) \rightarrow \infty$ since $\partial u/\partial y = 0$ at $y=0$ which is unrealistic.

\paragraph{The Carreau model:} Choosing the Carreau model for the viscosity (see Table~\ref{nonnewmodels}), the equations of the flow (\ref{eqn:cont_lub_cart}) and (\ref{eqn:momentum2_lub_cart}) are coupled with \eqref{eqn:momentum1_lub_cart}, which takes the form
\begin{equation}\label{eqn:momentum2_carreau}
\frac{\partial p}{\partial x} = \frac{\partial }{\partial y} \left[ \eta_\infty + \left(\eta_0 - \eta_\infty \right)    \left( 1+ \lambda^2 \abs{\frac{\partial u}{\partial y}}^2 \right)^{\frac{n_c-1}{2}} \abs{\frac{\partial u}{\partial  y}}  \right].
\end{equation}
This expression cannot be integrated analytically for $u(y)$, so we solve it numerically via the in-built \texttt{bvp5c} function in \textsc{Matlab}.

\paragraph{The Ellis model:} Choosing the Ellis model from Table~\ref{nonnewmodels}, the velocity of the fluid can be expressed analytically as
\begin{equation}\label{eqn:velocityellis}
u(y) = \frac{1}{\eta_0}\frac{\partial p}{\partial x} \left[ \frac{R^2 - \abs{y}^2}{2} + \left(\frac{1}{\beta}\frac{\partial p}{\partial x} \right)^{\alpha -1} \frac{R^{\alpha +1} - \abs{y}^{\alpha +1}}{\alpha +1} \right] \, .
\end{equation}

In order to compare the four models and their velocity fields we take typical parameter values from the literature (Table~\ref{Table:PhysicalParameters_flow}). The viscosity depends on temperature, so for the whole paper we use parameters consistent with a typical body temperature, 37ºC. We assume a micro vessel with a radius  $R=20$\,$\mu$m and the volume flux is $Q=8\times10^{-13}$\,m$^3$/s, consistent with average velocity measurements in capillaries or small arterioles \cite{Wan16}. For our simulations, we will use a reference length several times larger than the radius, $L=500$\,$\mu$m.

\begin{figure}[h]
	\centering
	\subfigure[]{\includegraphics[width=0.49\textwidth]{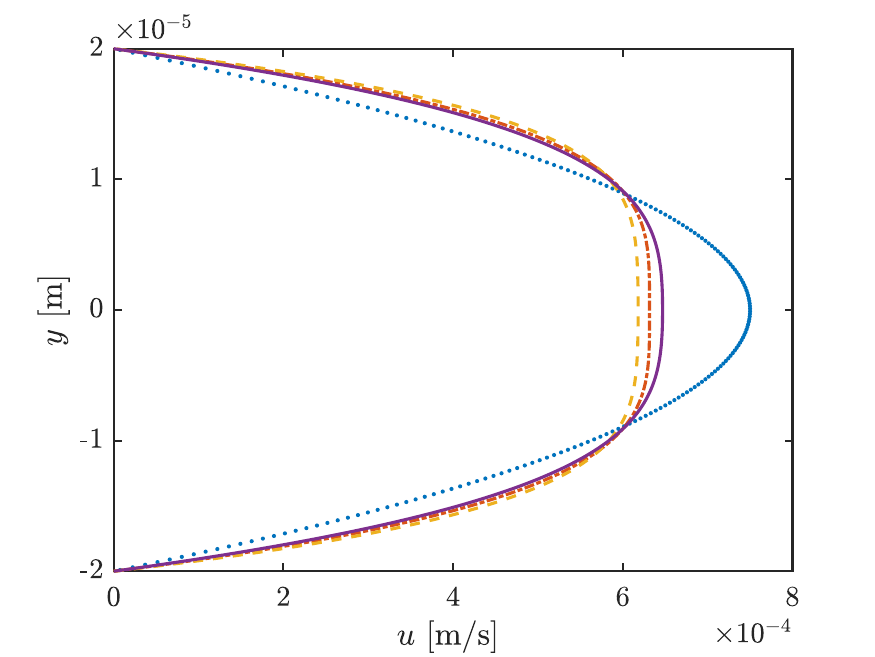}}
	\subfigure[]{\includegraphics[width=0.49\textwidth]{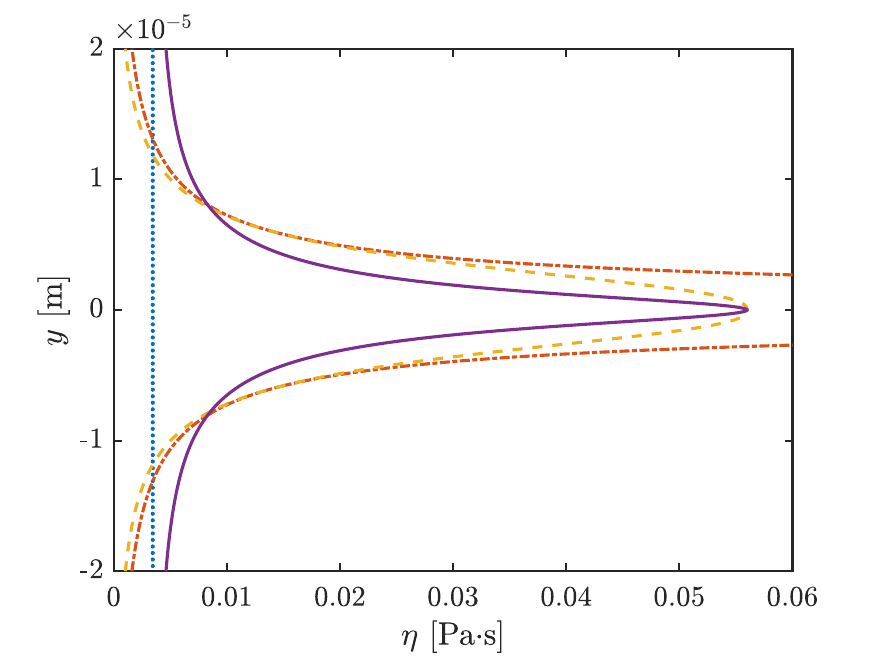}}
	\caption{Comparison of (a) velocity $u(y)$ and (b) viscosity $\eta(y)$ profiles for the Newtonian (dotted line), power-law (dashed-dotted line), Carreau (solid line) and Ellis model (dashed line).}
	\label{fig:comp_velocity}
\end{figure}
The velocity  and  corresponding viscosity profiles for blood, obtained from the four different models, are compared in Figures \ref{fig:comp_velocity}(a) and \ref{fig:comp_velocity}(b), respectively. We observe that although the difference in velocity profiles is small for all models, the corresponding viscosities can differ significantly. The Ellis and Carreau models show good agreement in both velocity and viscosity. In the literature, the Carreau model is generally preferred due to its ability to predict both Newtonian plateaus. However for \emph{in vivo } blood flow it is unlikely that the high shear plateau will be reached suggesting that the Ellis model provides an adequate alternative approximation. This is an important observation, since the Ellis model yields an analytical expression for the flow. This not only provides a better understanding of the factors affecting the flow but also considerably simplifies the numerical study. In Section \ref{numsec} we will verify that the Carreau and Ellis models provide similar results for the nanoparticle distribution. The Newtonian model has a constant viscosity and the power-law model cannot predict the Newtonian plateaus. Given that the motion of the magnetic particles is dependent on the viscosity of the fluid it travels through it is clear that both Newtonian and power-law models are not appropriate for modelling blood.

\section{Nanoparticle motion in a non-Newtonian flow subject to an external magnetic field}
\label{sec:conceq}

The behaviour of the concentration of magnetic nanoparticles in the bloodstream is obtained following the continuum model developed by Grief and Richardson \cite{grief2005mathematical}. The governing equation describing the motion of magnetic particles in the blood stream is an advection-diffusion equation for the particle concentration $c(x,y,t)$:
\begin{equation}\label{eqn:diff_cart}
\frac{\partial c}{\partial  t} + \nabla \cdot \left[(\textbf{u}_F + \textbf{u}_p) c \right]= \nabla \cdot\textbf{J}_{\mathrm {diff}}\,,
\end{equation}
where $\textbf{u}_F$ is the fluid velocity (as discussed in the previous section), $\textbf{u}_p$ is the particle velocity and $\textbf{J}_{\mathrm {diff}} = - D\,\nabla c$ is the diffusion flux.

The diffusion is due to Brownian motion and shear-induced diffusion and hence $D = D_{\mathrm {Br}} + D_{\mathrm {sh}}$. Shear-induced diffusion arises due to the fact that the RBCs suspended in plasma collide with each other causing random motion with a diffusive character. As recently demonstrated by Liu and coworkers \cite{liu2019nanoparticle, liu2018nanoparticle} via a lattice-Boltzmann multiscale simulations, the diffusion due to the Brownian motion in the case of nanoparticle transport in a small vessel can be important and, in some cases, even the predominant diffusion process. In particular, for nanoparticles smaller than 100\,nm they found that $D_{\mathrm {Br}}/D_{\mathrm {sh}}\gg1$, that is Brownian diffusion is dominant for particles in this size range \cite{liu2018nanoparticle}. Using the Stokes-Einstein equation for the diffusion of spherical particles through a shear thinning fluid \cite{fournier2017basic}, we can write the Brownian diffusion coefficient as
\begin{equation} \label{eqn:browniandiff}
D_{\mathrm {Br}} = \frac{k_\mathrm {B} T}{6 \pi \, \eta(\dot{\gamma}) \, a},
\end{equation}
where $k_\mathrm {B}$ is the Boltzmann constant, $T$ is the absolute temperature and $a$ is the particle radius. On the other hand, the shear-induced diffusion contribution can be approximated by
\begin{equation} \label{eqn:diffshear}
D_{\mathrm {sh}} = K_{\mathrm {sh}} \left(r_{\mathrm {RBC}}\right)^2 \dot{\gamma},
\end{equation}
where $K_{\mathrm {sh}}$ is a dimensionless coefficient that depends on the blood cell concentrations and $r_{\mathrm {RBC}}$ is the red blood cell radius. The coefficient $K_{\mathrm {sh}}$ is difficult to measure but the value used in Table \ref{Table:PhysicalParameters_conc} is considered representative in the literature \cite{grief2005mathematical}. Hence, 
\begin{equation} \label{eqn:diffflux}
\textbf{J}_{\mathrm {diff}} = - D \nabla c =
- \left( \frac{k_\mathrm {B} T}{6 \pi \, \eta(\dot{\gamma}) \, a} + K_{\mathrm {sh}} \left(r_{\mathrm {RBC}}\right)^2 \dot{\gamma} \right) \nabla c.
\end{equation}
The particle velocity is found by balancing hydrodynamic and magnetic forces. Using the definition of the Stokes drag, representing the hydrodynamic force on a spherical particle of radius $a$ moving through a viscous fluid, we have
\begin{equation}\label{eqn:stokesforce}
\textbf{F}_{\mathrm {St}} = 6 \pi a \,\eta(\dot{\gamma}) \textbf{u}_p \, .
\end{equation}
The particles reach equilibrium velocity when $\textbf{F}_{\mathrm {St}}$ balances the magnetic force $\textbf{F}_{\mathrm {mag}}$ and this leads us to the expression
\begin{equation}\label{eqn:velocitymagneticparticle}
\textbf{u}_p = \frac{ \textbf{F}_{\mathrm {mag}}}{6 \pi a \,\eta(\dot{\gamma})}\,.
\end{equation}
In order to exert a magnetic force on magnetic nanoparticles in the vessels, a magnetic field gradient is required at a distance. We can define the magnetic force $\mathbf{F}_{\mathrm {mag}}$ on a single particle in a magnetic field $\mathbf{B}$ as
\begin{equation}\label{eqn:magneticforcegeneral}
\mathbf{F}_{\mathrm {mag}} = (\mathbf{m} \cdot \nabla) \mathbf{B},
\end{equation}
Nacev \textit{et al.} \cite{nacev2011behaviors} have shown that for a magnet held at a long distance compared to the width of the vessel, we can assume the magnetic force is approximately constant in the vertical direction which avoids the need to solve Maxwell's equations. Richardson \textit{et al.} \cite{richardson2010particle}, for example, use a constant value for the magnetic force given by
\begin{equation}\label{eqn:F0}
F_0 = \frac{4}{3} \pi a^3 \rho \mathbf{M} \Upsilon B_g,
\end{equation}
where $\Upsilon$ is the magnetite volume fraction and $B_g$ is the gradient magnetic field. This is what we assume in this work. We consider a constant magnetic force acting perpendicular to the flow, i.e. $\mathbf{F}_{\mathrm {mag}} = F_0\,\mathbf{j}$. Therefore, in combination with \eqref{eqn:velocitymagneticparticle} we obtain that $\textbf{u}_p = (0, v_p(y))$ where $v_p =-F_0/(6 \pi a \,\eta(\dot{\gamma}))$. Finally, since the vertical fluid velocity is negligible, $\textbf{u}_{\text{tot}} = (u_F(y), v_p(y))$ and $D = D(y)$, the concentration equation (\ref{eqn:diff_cart}) becomes
\begin{equation}\label{eqn:advdiff_final}
\frac{\partial c}{\partial  t} + u_F \frac{\partial c}{\partial x }+  \frac{\partial(v_p c)}{\partial y} =   D \frac{\partial^2 c}{\partial x^2} + \frac{\partial }{\partial y}\left( D \frac{\partial c}{\partial y}\right).
\end{equation}

We assume that the particles can flow out of the vessel through the walls with a certain vascular permeability $\kappa$. This results in Robin boundary conditions at the top and the bottom of the channel, of the form
\begin{equation}\label{eqn:bcswall}
\left(\textbf{u}_{\text{tot}} c - D \nabla c \right)\cdot \mathbf{n}= \kappa \, c \qquad \text{on}\qquad  y=\pm R\,,
\end{equation}
where $\mathbf{n}$ is the unit outward normal vector at the vessel walls. 

As  predicted by Richardson \textit{et al.} \cite{richardson2010particle} via  matched asymptotic expansions, the parameter $\kappa$ plays an important role in determining the position of the particles deposited onto the vessel wall. In fact, their outer solution shows how small values of the permeability are responsible for the formation of a boundary layer region in the immediate vicinity of the wall where the advective flux balances the diffusive flux and the thickness of the vessel wall prevents particles from flowing out and this may hamper the proposed treatment. In this work, a reference value of $\kappa = \mathcal{O}(10^{-6})$ is chosen \cite{Lim20}.

Assuming that particles enter the channel due to an injection of 3\,s duration in the vicinity of $x=0$, we have 
\begin{align}
u_F c_{\mathrm{in}}(y,t) = \left( u_F c  - D \frac{\partial c}{\partial  x}\right) \qquad \text{on}\qquad  x=0\,,
\end{align}
where the concentration $c_\mathrm{in}(y,t)$ describes an injection in the central region of the vessel (see Appendix \ref{injection}). 

At the channel outlet 
\begin{equation}\label{eqn:bcoutflow}
\left( u_F c  - D \frac{\partial c}{\partial  x}\right)=0
\qquad \text{on}\qquad  x=L\,.
\end{equation}
However, this outlet boundary condition is irrelevant since particles never reach the outlet of the vessel in our simulations. The initial particle concentration in the channel is zero, hence $c(x,y,0) = 0$.

\begin{table}[h]
	\centering
	\begin{tabular}{lcccc}
		\hline
		\noalign{\vskip 2mm}
		\textbf{Quantity} & \textbf{Symbol} & \textbf{Value} & \textbf{Units}  & \textbf{References} \\
		\noalign{\vskip 2mm}
		\hline
		\noalign{\vskip 2mm}
		Blood cell radius & $r_{\mathrm{RBC}}$ & $4.2 \times 10^{-6}$ & m  & \cite{grief2005mathematical} \\
		Shear diffusion coef. & $K_{\mathrm{sh}}$ & $ 5 \times 10^{-2}$ & No.  &  \cite{grief2005mathematical} \\
		Reference concentration & $c_{0}$ & 1 & mol m$^{-3}$  & \cite{nacev2011behaviors} \\
		Particle radius & $a$ & $15 \times 10^{-9}$ & m & \cite{grief2005mathematical} \\
		Boltzmann constant & $k_\mathrm{B}$ & $1.38 \times 10^{-23}$ & m$^2$ kg s$^{-2}$ K$^{-1}$ & \cite{fournier2017basic} \\
		Temperature &  $T$ & 310.15 & K & \cite{fournier2017basic} \\
		\noalign{\vskip 2mm}
		\hline
	\end{tabular}
	\caption{Parameter values chosen for the advection-diffusion equation.}
	\label{Table:PhysicalParameters_conc}
\end{table}

\subsection{Simulation of the nanoparticle concentration}\label{numsec}

To solve \eqref{eqn:advdiff_final} numerically, we first nondimensionalise the equation (see Appendix \ref{subs:nondimdiffusion}). Then, we use an explicit Euler scheme in time, with first order upwind approximations for the advection terms and central differences for the second order derivatives (see Appendix \ref{subs:numericalscheme} for more details on the numerical scheme). The velocity $u_F$ is presented in Section \ref{subsec:comparison}, where the Newtonian and Ellis fluid models yield analytical expressions for $u_F$ while the Carreau model velocity is determined numerically. The vertical particle velocity $v_p$ is calculated via equation (\ref{eqn:velocitymagneticparticle}), with a constant value of the magnetic force $F_0$. Advection terms, as usual, will dominate where the flow rate is non-negligible but the diffusive contribution becomes important near the wall of the vessel, where the fluid has nearly zero velocity (see discussion on boundary layers in Appendix \ref{subs:nondimdiffusion}). For the simulations shown in the next section, we use a square unit domain with a resolution of 400 nodes in the $x$ and $y$ directions. The time step, $\Delta t$, is suitably chosen to satisfy the stability requirements of the scheme.

At the beginning of the process, there are no particles in the vessel, that is $c(x,y,0)=0$, but at $t=0$ we inject a concentration of particles $c_{\mathrm{in}}(y,t)$ at the central region of the boundary $x=0$. To avoid the numerical difficulty associated with an abrupt change in concentration at this boundary, we assume that the concentration $c_{\mathrm{in}}(y,t)$ progressively increases with time until it reaches a maximum value, $c_0$, at $t=3$\,s (see Appendix \ref{injection}).  

\subsection{Mesh convergence}

Here we provide results of a spatial and temporal convergence study before presenting numerical predictions based on the numerical scheme given in Appendix \ref{subs:numericalscheme}. Following the notation from Appendix \ref{subs:numericalscheme}, all variables used in this subsection are dimensionless. The reference numerical approximation, $c^{\text{ref}}$, is generated on a fine mesh with $n_x=400$ and $n_y=400$ grid points in the $x$ and $y$ directions, respectively, with $n_x \Delta x=n_y \Delta y =1$,  and time step $\Delta t_4$ where $\Delta t_k = \Delta t_0/2^{k-1}$, $k=1,\ldots,4$,  and $\Delta t_0 = 0.2\times 10^{-5}$. Let $M_n$ denote the mesh with $(25 \times 2^n)^2$ cells with $n=1,\ldots,4$. 

To investigate convergence with respect to mesh size, the $l^2$-norm of the difference between the numerical approximations on meshes $M_4$ and $M_n$ is computed for $n=1,\ldots,3$ at the end of the simulation $t=0.8$. This is denoted by $E_n$. Since the concentration at the lower boundary $y=-1/2$ is important in our study, we also compute the $l^2$-norm at this boundary denoted by $E_n^{lb}$. These measures of the error are tabulated in Table \ref{En}. They confirm second-order convergence of the numerical approximation with respect to mesh spacing as shown in the final column where the rate of convergence $p_n$ is computed using $p_n=\log_2(E_{n/2}/E_{n})$.

\begin{figure}[h]
	\centering
    \includegraphics[width=0.5\textwidth]{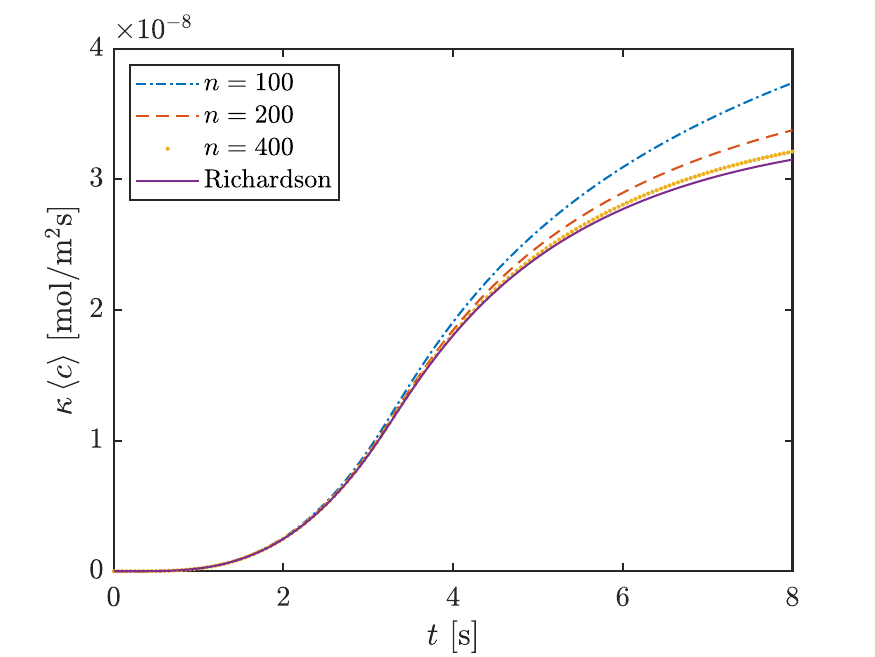}
	\caption{Influence of mesh size on the evolution of the average nanoparticle flux through the boundary $y=-R$ for the Ellis model with magnetic force $F_0 = 0.5\times10^{-14}$\,N. Also shown is an application of Richardson extrapolation. }
\label{F0_error}
\end{figure}

The influence of mesh size on the evolution of the average nanoparticle flux through the bottom boundary is shown in Figure \ref{F0_error} in dimensional units for the Ellis model with magnetic force $F_0 = 0.5\times10^{-14}$\,N. Also shown in Figure \ref{F0_error} is an application of Richardson extrapolation based on extrapolating the approximations computed on meshes $M_2$, $M_3$ and $M_4$. This demonstrates convergence of this quantity with mesh refinement with enhanced accuracy obtained through use of Richardson extrapolation.

Next we investigate convergence with respect to time step. In this temporal convergence study we use the finest mesh $M_4$ and compute the $l^2$-norm of the difference between the numerical approximations obtained using time steps $\Delta t_4$ and $\Delta t_k$ for $k=1,\ldots,3$ at time $t=0.8$. These measures of the error are tabulated in Table \ref{Ek}. They confirm first-order convergence of the numerical approximation with respect to time step as shown in the final column where the rate of convergence $p_k$ is computed using $p_k=\log_2(E_{k}/E_{k-1})$ .

\begin{table}[h]
\centering
\begin{tabular}{|c|c|c|c|c|}
\hline
 $n_x=n_y$ &  $\Delta x=\Delta y$ & $E_{n}$ & $E_{n}^{lb}$ & $p_n$  \\
 \hline
 50   &  0.02  &  0.0176 & 0.0182 & - \\
 100  &  0.01  &  0.0052 & 0.0051 & 1.7577 \\
 200  & 0.005  &  0.0012 & 0.0012 & 2.1621 \\
\hline
\end{tabular}
\caption{Convergence analysis with respect to mesh size.} 
\label{En}
\end{table}

\begin{table}[h]
\centering
\begin{tabular}{|c|c|c|c|c|c|}
\hline
$k$ & $\Delta t_k$ & $E_{k}$ & $E_{k}^{lb}$ & $p_k$ \\
 \hline
1 & 0.2 $\cdot10^{-5}$  &  5.6478$\cdot10^{-7}$ &  2.6025$\cdot10^{-7}$ & - \\
2 & 0.1 $\cdot10^{-5}$  &  2.8232$\cdot10^{-7}$ &  1.3011$\cdot10^{-7}$ & 1.0004 \\
3 & 0.05 $\cdot10^{-5}$ &  1.4114$\cdot10^{-7}$ &  6.5055$\cdot10^{-8}$ & 1.0002 \\
\hline
\end{tabular}
\caption{Convergence analysis with respect to time step.} 
\label{Ek}
\end{table}

\section{Results and discussion}\label{sec:results}

We first analyse the effects that the different fluid models have on the motion of the particles in the bloodstream. As demonstrated in Section~\ref{sec:bloodnonnewt} the power-law model gives unrealistic values for the viscosity and therefore the results for this case will not be discussed. The results presented here, correspond to numerical simulations of equation \eqref{eqn:advdiff_final} applying an injection for three seconds at $x=0$. 

\subsection{Nanoparticle migration in Newtonian blood flows }

\begin{figure*}
	\centering
    \includegraphics[width=0.7\textwidth]{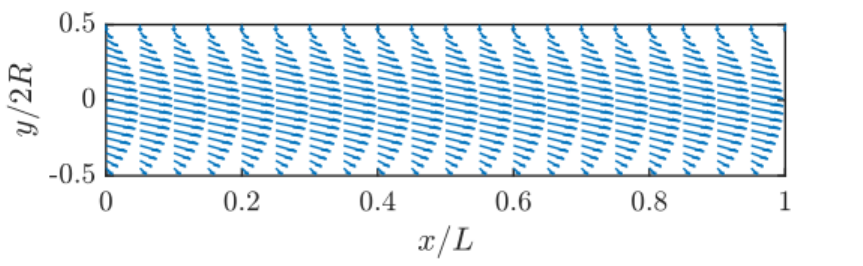}
    \includegraphics[width=0.8\textwidth]{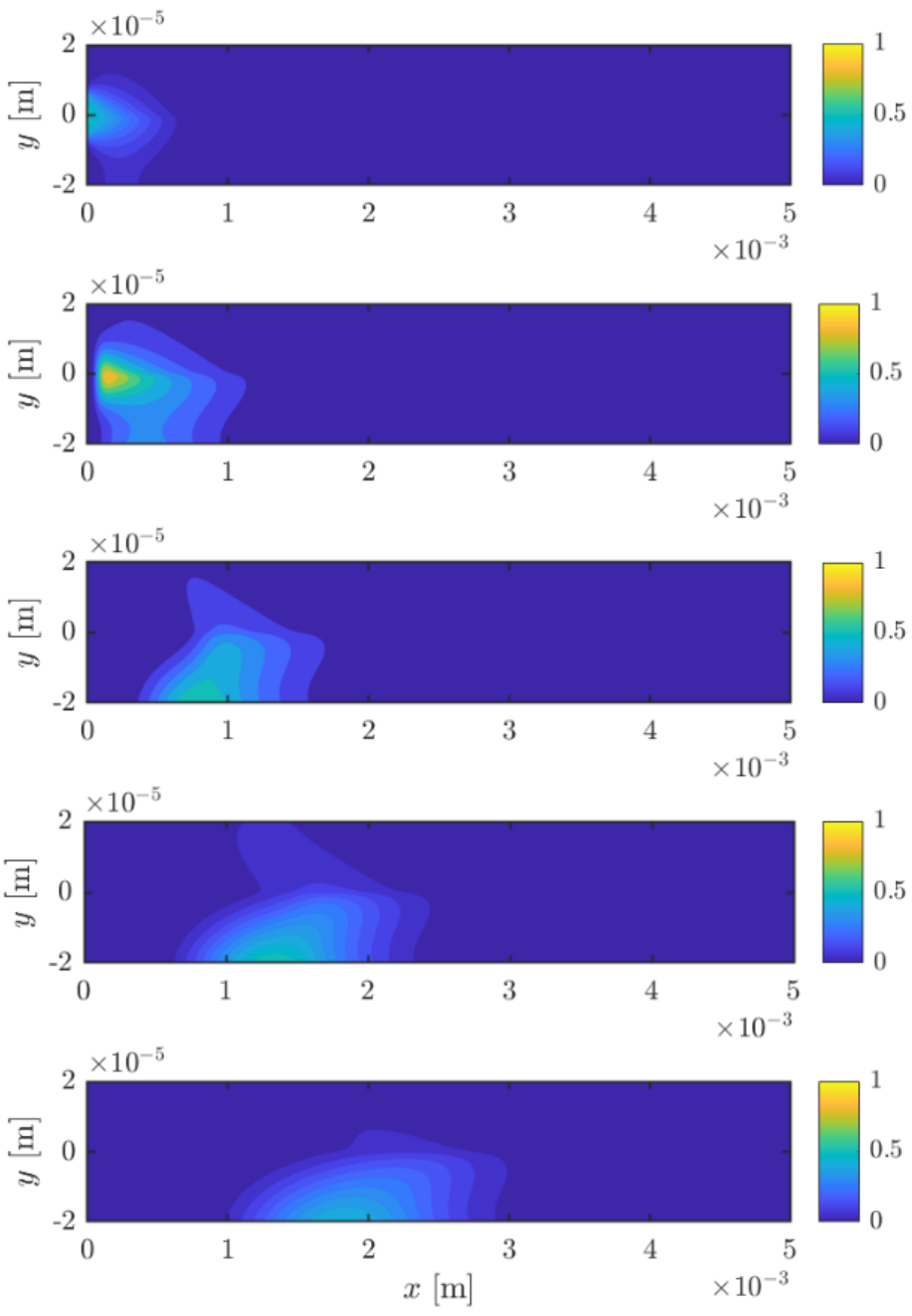}
	\caption{Snapshots of the concentration of magnetic nanoparticles, $c/c_0$, in a capillary vessel at five different times ($t=1.6$s, $t=3.2$s, $t=4.8$s, $t=6.4$s, $t=8$s), using the Newtonian model for blood flow and with a constant magnetic force equal to $F_0 = 0.5 \times 10^{-14}$ N. The top panel represents the velocity field (equation \eqref{eqn:velocitynewtonian}). Other parameter values as in Table~\ref{Table:PhysicalParameters_flow}.}
	\label{fig:Newtonian_F1-13}
\end{figure*}

Figure \ref{fig:Newtonian_F1-13} shows the evolution of the concentration of nanoparticles in the vessel under the influence of a  magnetic force $F_0 = 0.5 \times 10^{-14}$\,N when the Newtonian approximation for blood is considered. The panel on top represents the velocity field, $\boldsymbol{u}_{\text{tot}}$. Since the velocity of the particles depends on the viscosity of the fluid and the Newtonian flow assumes a constant value, $\mu$ then $v_p=F_0/6\pi a \mu$, the changes in the velocity field are only due to the parabolic profile $u_\mathrm{F}(y)$ and only in the $y$-direction. Furthermore, we observe how the particles near the vessel wall experience a much smaller drag force with respect to those at the center of the vessel (i.e., the horizontal component of the velocity is smaller near the wall than at the center) and, therefore, will react strongly to the magnetic force and deviate from the typical parabolic behaviour of the fluid. The colour maps show snapshots of the concentration of particles at five different times. We observe that the particles entering the bloodstream from the vessel inlet are immediately driven to the lower wall of the vessel, where some of them will permeate the vessel wall. At $t=6.4$\,s and $t=8$\,s, we observe the formation of a boundary layer where particles accumulate in the immediate proximity of the wall (e.g., at $t=8$\,s large values of $c$ appear near $x=0.2\cdot10^{-3}$\,m, at the lower boundary of the vessel), as theoretically predicted in \cite{richardson2010particle}.

\subsection{Nanoparticle migration in non-Newtonian blood flows }

\begin{figure*}
	\centering
    \includegraphics[width=0.7\textwidth]{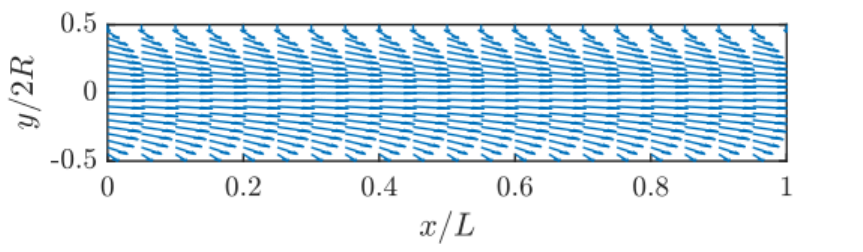}
    \includegraphics[width=0.8\textwidth]{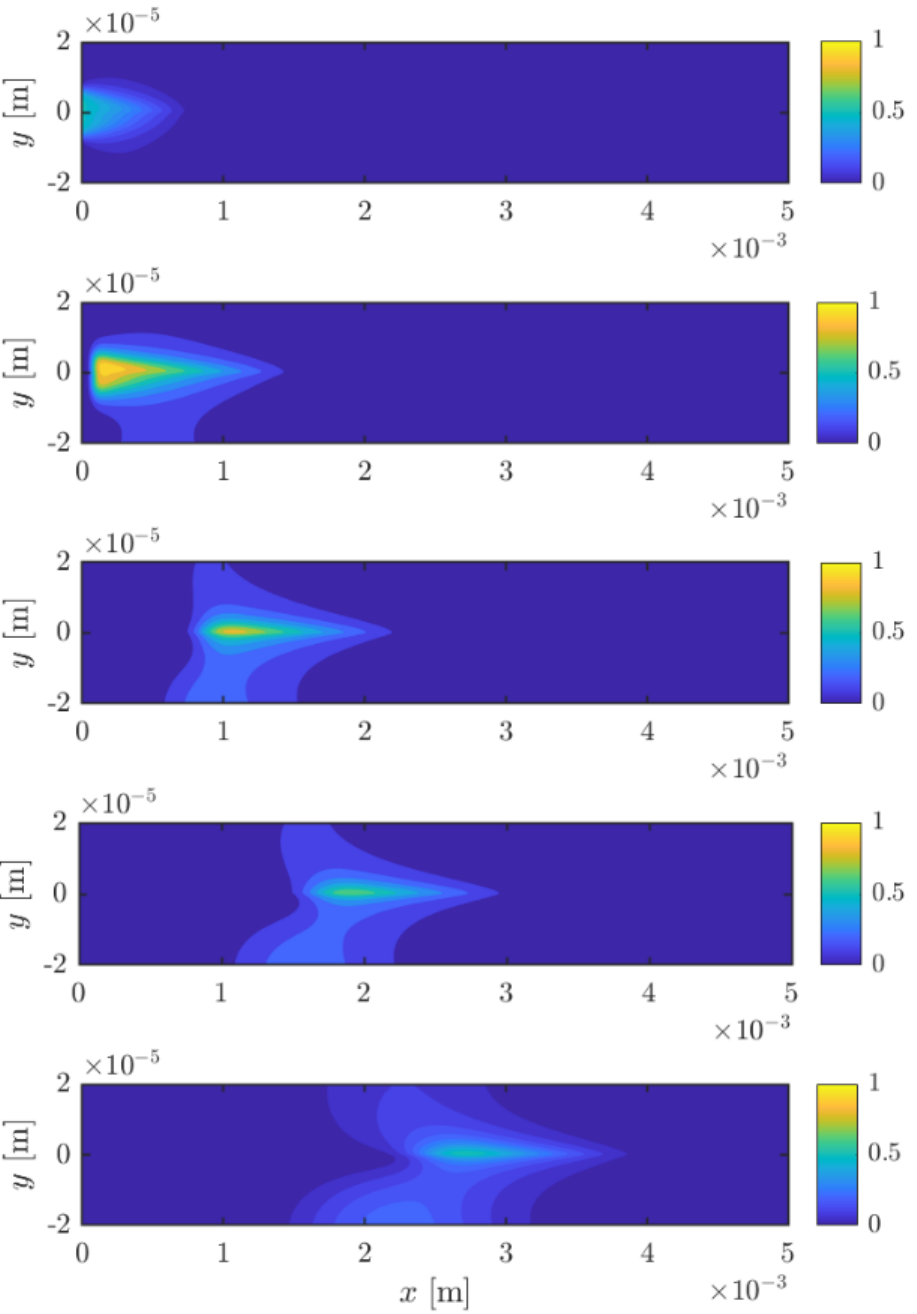}
	\caption{Snapshots of the concentration of magnetic nanoparticles, $c/c_0$, in a capillary vessel at five different times ($t=1.6$s, $t=3.2$s, $t=4.8$s, $t=6.4$s, $t=8$s), using the Carreau model for blood flow and with a constant magnetic force equal to $F_0 = 0.5 \times 10^{-14}$ N. The top panel represents the velocity field (see eq. \eqref{eqn:momentum2_carreau}). Other parameter values as in Table~\ref{Table:PhysicalParameters_flow}. }
	\label{fig:Carreau_F1-13}
\end{figure*}

\begin{figure*}
	\centering
    \includegraphics[width=0.7\textwidth]{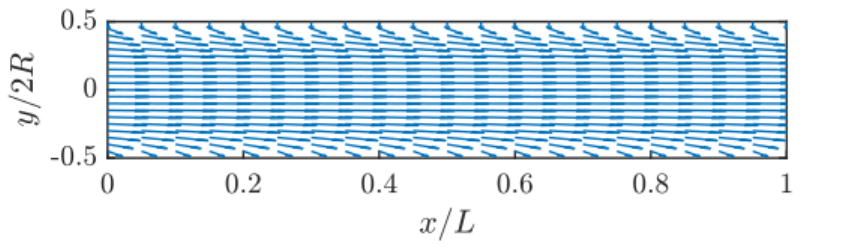}
    \includegraphics[width=0.8\textwidth]{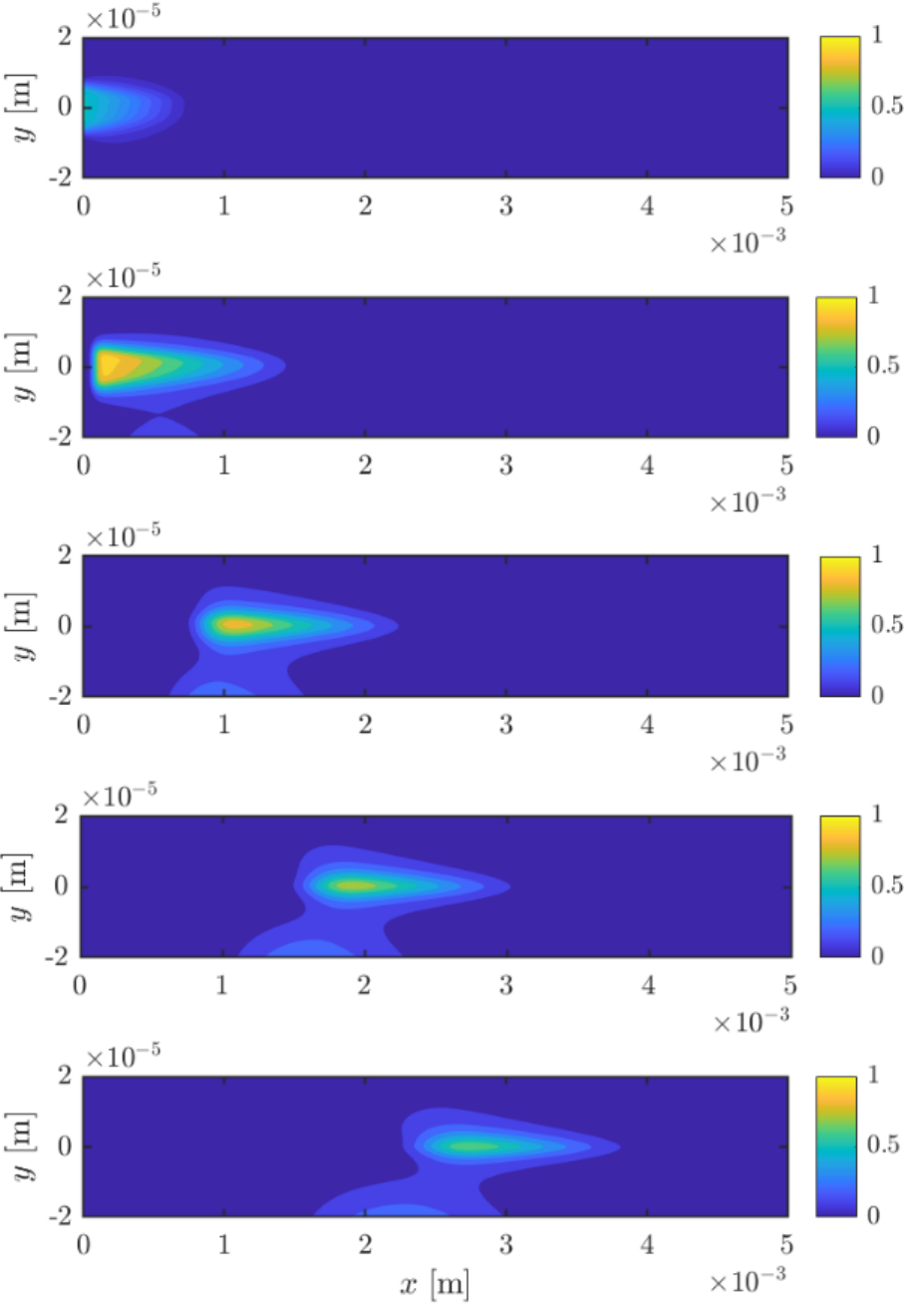}
	\caption{Snapshots of the concentration of magnetic nanoparticles, $c/c_0$, in a capillary vessel at five different times ($t=1.6$s, $t=3.2$s, $t=4.8$s, $t=6.4$s, $t=8$s), using the Ellis model for blood flow and with a constant magnetic force equal to $F_0 = 0.5 \times 10^{-14}$ N. The top panel represents the velocity field (equation \eqref{eqn:velocityellis}). Other parameter values as in Table~\ref{Table:PhysicalParameters_flow}.}
	\label{fig:Ellis_F1-13}
\end{figure*}

In Figures \ref{fig:Carreau_F1-13} and \ref{fig:Ellis_F1-13} the analogous plots to those of Figure \ref{fig:Newtonian_F1-13} are shown for the Carreau and the Ellis models, respectively. As expected, both models yield very similar particle concentrations. However, in contrast to the Newtonian case, not all the nanoparticles are forced out at the lower boundary. This change in behaviour may be attributed to the high viscosity values characterising the non-Newtonian models in the vicinity of $y=0$ (see Figure~\ref{fig:comp_velocity}b), due to which the particles find it difficult to escape this `central' region. Since the vertical velocity is proportional to $1/\eta$, in the Newtonian model the vertical velocity is significant and much larger than the corresponding one for the Ellis and Carreau models where the higher viscosity values lead to  significantly lower vertical velocities in the `central' region. Near the boundaries the viscosity of the Ellis and Carreau models takes a similar value to the Newtonian fluid and, hence, so does the vertical velocity. This may be observed from the velocity profiles shown in the top panels. The differences found in the concentration maps from Fig.~\ref{fig:Newtonian_F1-13}-\ref{fig:Ellis_F1-13} are mainly caused by the changes in the viscosity between models and clearly demonstrate that a Newtonian fluid model severely overestimates the effect of the magnetic force on the nanoparticles. 

Note the advection term in \eqref{eqn:diff_cart} is the dominant mechanism for nanoparticle transport in all fluid models studied (see discussion in Appendix~\ref{subs:nondimdiffusion}), but diffusion also contributes to the observed reduction in concentration, as particles spread out. This can be observed in all three cases (Figures \ref{fig:Newtonian_F1-13}-\ref{fig:Ellis_F1-13}) where the colored central region increases size with time. Actually, diffusion allows some reduced number of nanoparticles reach the upper vessel wall (see, for instance, panels for $t= 6.4$\,s in Figures~\ref{fig:Newtonian_F1-13}-\ref{fig:Carreau_F1-13}). 

\subsection{Varying the strength of the magnetic field}
\begin{figure*}
	\centering
    \includegraphics[width=0.6\textwidth]{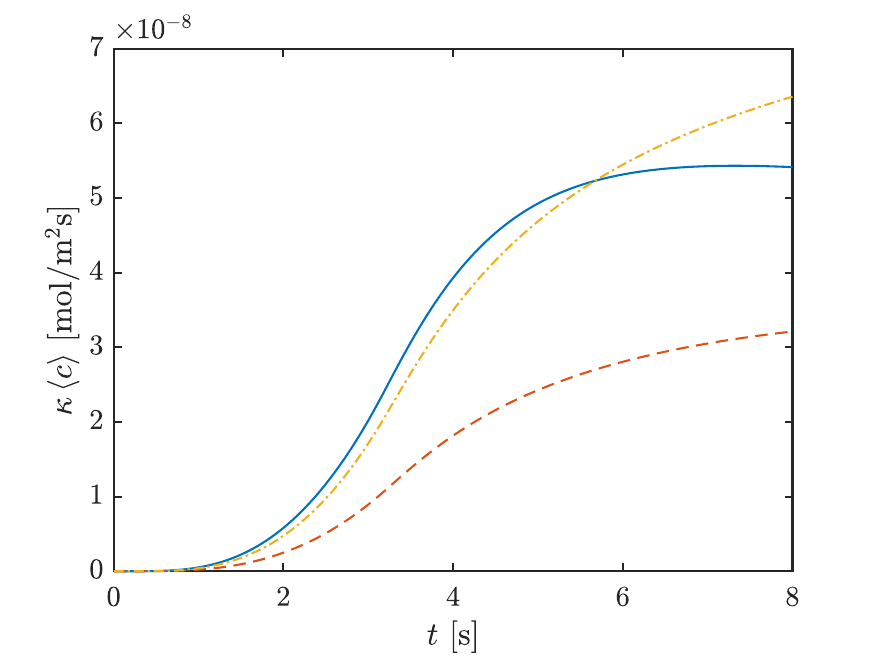}
	\caption{The average particle flux at the lower wall for the Newtonian and Ellis models versus time with magnetic force $F_0 = 0.5 \times 10^{-14}$ N (solid and dashed lines, respectively), and the Ellis model for a doubled force, $F_0 = 1 \times 10^{-14}$ N (dash-dotted line).	}
	\label{f0_vs_F0}
\end{figure*}

Having demonstrated that the non-Newtonian models are more realistic, we now aim to find a value of the magnetic force $F_0$ capable of driving the same number of particles to the lower wall of the vessel as the Newtonian fluid, for the same capillary vessel length and radius as in previous simulations. For doing so, we analyse the average particle flux crossing the lower vessel wall as a function of time, which is computed using $c\vert_{y=-R}$ via
\begin{equation}\label{avF}
\kappa\, \langle c\rangle = \frac{\kappa}{L}\int_{0}^{L} c\vert_{y=-R}\, dx\,.
\end{equation}
Since the Ellis and Carreau models provide similar results, for simplicity we use  the Ellis model as a representative non-Newtonian fluid. In Figure \ref{f0_vs_F0} we present the evolution of the average particle flux \eqref{avF} with time for the Newtonian (solid line) and Ellis (dashed line) fluids taking $F_0 = 0.5 \times 10^{-14}$\,N. We also simulate a third case corresponding to the Ellis fluid (dash-dotted line) with a magnetic force $F_0 = 1 \times 10^{-14}$\,N, i.e., a magnetic force two times larger. As expected, the average flux for the Ellis fluid with $F_0 = 0.5 \times 10^{-14}$\,N is much lower than the Newtonian fluid for the same magnetic force. For instance, at $t=8$\,s the average flux for the Ellis fluid is a 41\% lower than the Newtonian one. However, by doubling the strength of the magnetic field, we observe that the flux for the Ellis model gets closer to the Newtonian case with $F_0 = 0.5 \times 10^{-14}$\,N. To facilitate the comparison between these two cases, we numerically integrate the corresponding curves over the total simulation time, obtaining $2.4977\times10^{-7}$\,mol/m$^2$ and $2.4992\times10^{-7}$\,mol/m$^2$ for the Newtonian with $F_0 = 0.5 \times 10^{-14}$\,N and the Ellis with $F_0 = 1 \times 10^{-14}$\,N fluid, respectively. Hence, increasing the magnetic force to $F_0 = 1 \times 10^{-14}$\,N is enough to attract the same number of particles to the lower vessel wall in the chosen simulation time.

\begin{figure*}
	\centering
    \includegraphics[width=0.7\textwidth]{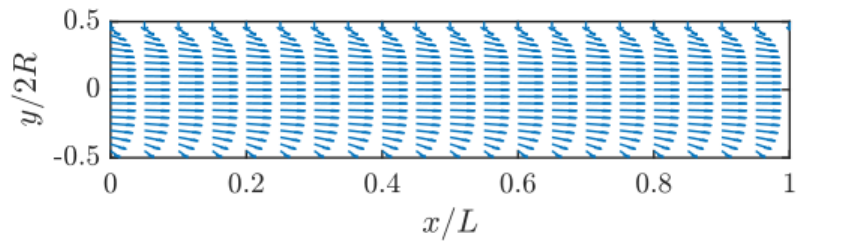}
    \includegraphics[width=0.8\textwidth]{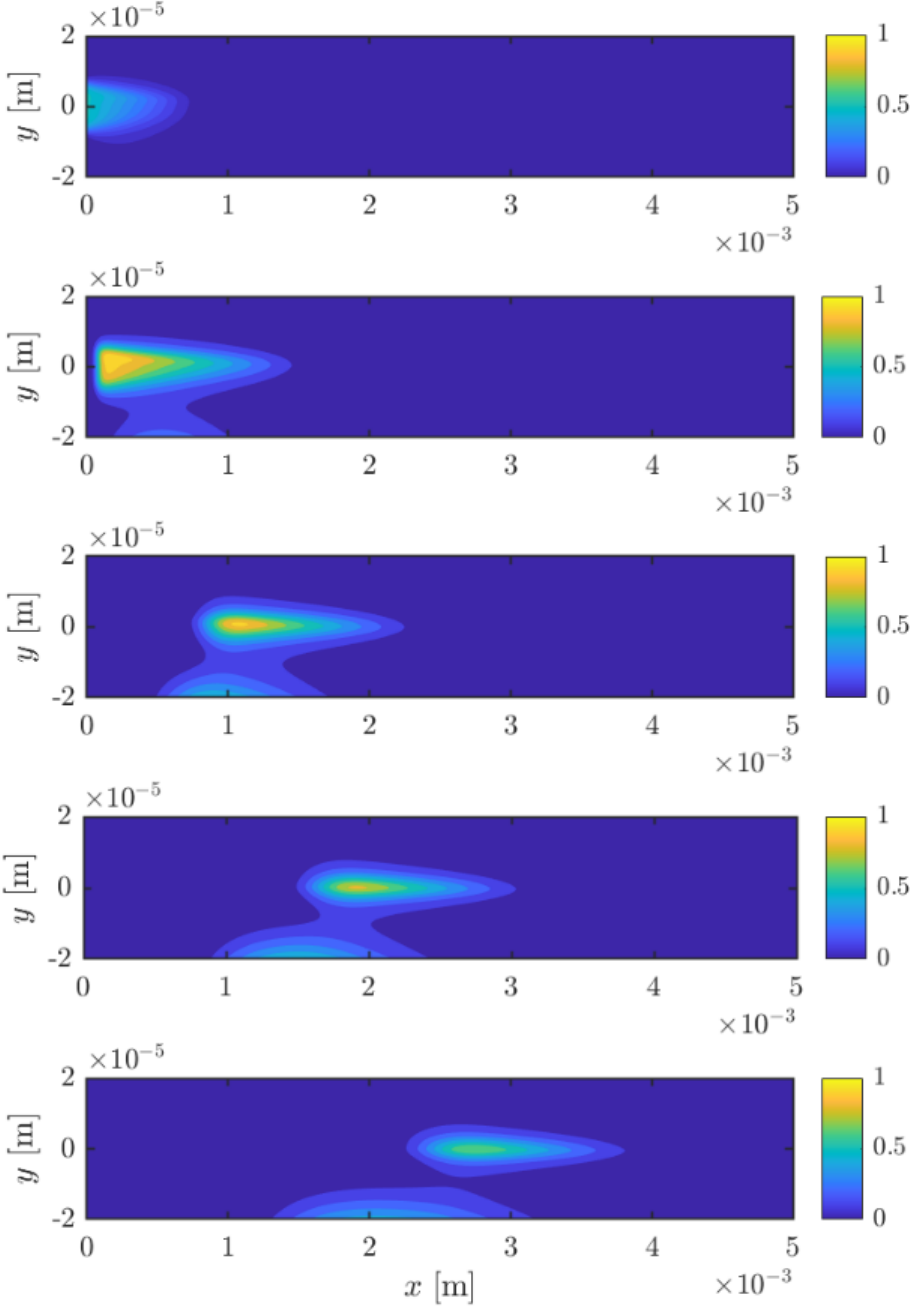}
	\caption{Snapshots of the concentration of magnetic nanoparticles, $c/c_0$, in a capillary vessel at five different times ($t=1.6$s, $t=3.2$s, $t=4.8$s, $t=6.4$s, $t=8$s),
	using the Ellis model for blood flow and with a constant magnetic force equal to $F_0 = 1 \times 10^{-14}$ N. The top panel represents the velocity field (equation \eqref{eqn:velocityellis}). Other parameter values as in Table~\ref{Table:PhysicalParameters_flow}.}
	\label{fig:Ellis_1F-14}
\end{figure*}

In Figure~\ref{fig:Ellis_1F-14}, we present the particle concentration maps for the Ellis model with $F_0 = 1 \times 10^{-14}$\,N. By comparing them with those in Figure \ref{fig:Ellis_F1-13}, we observe that the number of particles pushed to the lower boundary is clearly larger, consistent with the higher average flux of particles shown in Figure~\ref{f0_vs_F0}. Since increasing $F_0$ increases the vertical velocity, the formation of the boundary layer in the vicinity of the lower wall, where the y-advection term is dominant, is even clearer than in Figure~\ref{fig:Ellis_F1-13}.  

\subsection{Sensitivity under changes of the vessel wall permeability}

\begin{figure}[h]
	\centering
    \includegraphics[width=0.5\textwidth]{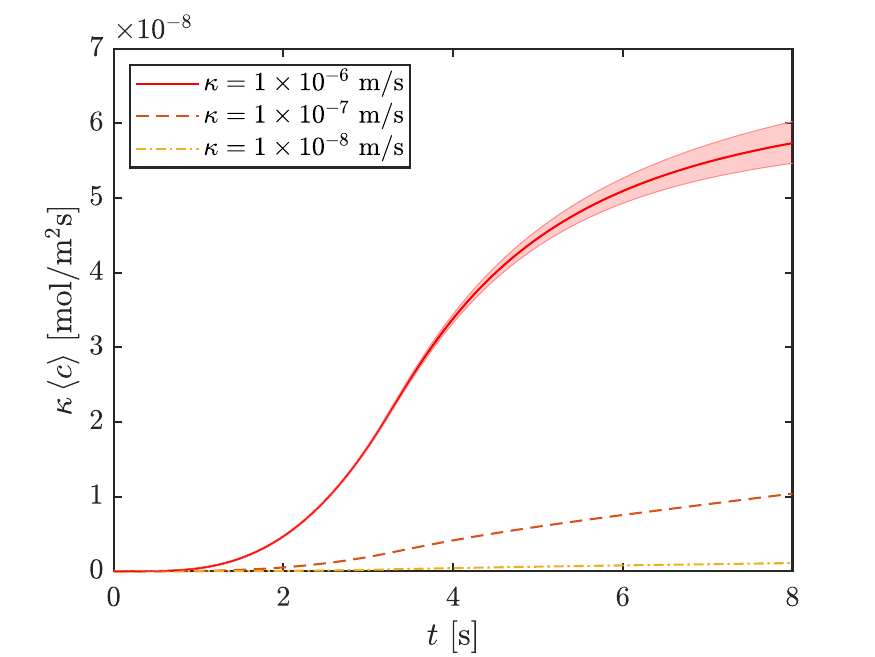}
	\caption{Evolution of the average nanoparticle flux through the boundary $y = -R$ as a function of time for  $F_0 = 1\times10^{-14}$\,N and three different permeabilities using the Ellis model. The shaded area represents the sensitivity of the nanoparticle flux for variations of the permeability within a 10\% around $\kappa = 1\cdot10^{-6}$\,m/s.  }
	\label{figure_change_kappa}
\end{figure}

We note that different tissues may have different permeabilities, and recent experiments indicate effective permeability values of the order of 10$^{-6}$\,m/s or smaller in tumors~\cite{Lim20}. As shown in Figure~\ref{figure_change_kappa}, a decrease in the permeability in our model results in a consistent decrease of the particle flux. Also shown in Figure~\ref{figure_change_kappa} is a sensitivity analysis with respect to the physically meaningful value of the permeability, i.e., $\kappa=1\cdot10^{-6}$ m/s. Increasing or decreasing this reference value of the permeability by 10\% with respect the reference value $\kappa = 1\cdot10^{-6}$\,m/s generates a change in the particle flux of around 5\% at the end of the simulation ($t=8$ s). This demonstrates that the results are robust with respect to small changes to the value of vascular permeability.

\section{Conclusions}
\label{sec:conclu}

We have formulated a model that describes the motion of magnetic nanoparticles in a blood vessel subject to an external magnetic field in order to optimize  magnetic drug targeting. The model consists of a system of nonlinear partial differential equations formed by the Navier-Stokes equations for the flow of blood and with an advection-diffusion equation for the concentration of nanoparticles. We consider Newtonian flow and three different non-Newtonian flows. Assuming a long two-dimensional vessel, the equations are significantly reduced and the system is then solved via analytical and numerical techniques.

We have accounted for all the forces involved in this physical process, combined with realistic choices for the parameter values. It has been shown that, in order to correctly simulate the delicate balance between hydrodynamic (Stokes drag) and magnetic forces in the vessel, it is crucial to choose an appropriate non-Newtonian model for blood. 

The first part of this paper examines the non-Newtonian behaviour of blood and the importance of choosing an appropriate model for the fluid viscosity. The Newtonian approximation proved to be inaccurate while the more commonly used power-law model exhibits an unbounded value for the viscosity at the centre of the vessel. The Carreau and Ellis models are both found to be good models for simulating blood behaviour, and they lead to very similar predictions for the velocity of the fluid. However, the Ellis model has a distinct advantage in that it permits an analytical expression for the fluid velocity, so we focus on it for this reason. 

In the second part magnetic nanoparticles were introduced in the flow. With such small particles both advection and diffusion effects can play an important role. The key result of the paper is that a Newtonian model predicts greater particle motion than an Ellis or Carreau model under the same external magnetic force. Specifically, for one case examined, a magnetic force had to be applied in the Ellis fluid that was twice as large than in the Newtonian fluid for the same number of particles to be absorbed through the vessel wall. These results show that models based on a Newtonian fluid can drastically overestimate the effect of the magnetic field and therefore we conclude that in order to accurately model nanoparticle drug targeting in realistic clinical situations the non-Newtonian behaviour of blood needs to be accounted for so that a sufficiently strong magnetic field is applied. 

In future work the model can be improved by imposing pulsatile flow, including the elasticity of the vessels due to the change in pressure and also by solving in more complex geometries. Furthermore, it can be combined with detailed experimental studies to optimize the delivery of drugs to specific regions.



\section*{Acknowledgments}

CF acknowledges financial support from the Spanish Ministry of Economy and Competitiveness, through the “Maria de Maeztu” Programme for Units of Excellence in R\&D (MDM-2014-0445). TM acknowledges the support of PID2020-115023RB-I00 financed by MCIN/AEI/
10.13039/501100011033/ and by “ERDF A way of making Europe”. FF acknowledges the support of the Serra-Hunter Programme of the Generalitat de Catalunya. TM acknowledges the support of the CERCA Programme of the Generalitat de Catalunya.



\section{Appendix A: Initial injection}
\label{injection}

\begin{figure*}
	\centering
    \includegraphics[width=0.6\textwidth]{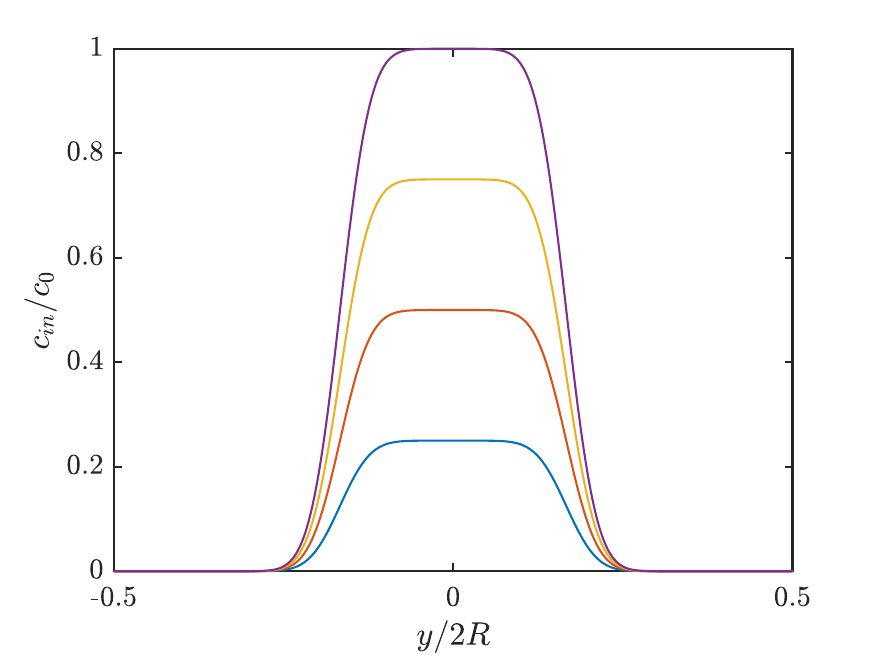}
	\caption{Profiles of $c_{\text{in}}$ at four different times (from bottom to top: $t=0.75$\,s, $t=1.5$\,s, $t=2.25$\,s and $t=3$\,s). In our simulations $M=20$. }
	\label{fig:inj}
\end{figure*}

To reproduce the injection of nanoparticles in the vessel we assume that the concentration of particles entering the vessel follow the distribution 
\begin{align}
c_{\text{in}} (y,t) = \frac{f(t)}{4}\,\text{erfc}\left[\frac{M}{2R}\left(y-\frac{R}{3}\right)\right]\left\{1+\text{erf}\left[\frac{M}{2R}\left(y+\frac{R}{3}\right)\right]\right\} c_0\,, 
\end{align}
which ensures that the injection is made at the central region of the channel, occupying approximately one third of the vessel diameter. The parameter $M$ controls the steepness of the error functions, and the time evolution is provided by
\begin{align}
f(t)=\begin{cases}
\frac{1}{3}\,t\qquad \text{if}\ 0\leq t\leq 3\,s\,,\\
0\qquad\ \ \text{otherwise\,,}
\end{cases}
\end{align}
allowing a progressive increase of particles with time, reaching the maximum $c_{0}$ at $t=3$\,s. The function $c_{\text{in}}$ satisfies the initial condition $c(x,y,0)=0$ and prevents jumps at the boundaries $y=\pm R$, since $c_{\text{in}}\rightarrow 0$ for $y\rightarrow\pm R$, thereby avoiding numerical instabilities. See profiles in figure \ref{fig:inj}. 

\section{Appendix B: Nondimensionalisation of the diffusion equation}
\label{subs:nondimdiffusion}

Using the non-dimensional variables (\ref{eqn:nondimvariab_cart}) and introducing new ones
\begin{equation}\label{eqn:nondimvariab2_cart}
t = \delta t \, \hat{t}, \qquad c = c_0 \, \hat{c}, \qquad u_F = U \hat{u}_F, \qquad v_p = W \hat{v}_p, \qquad D = \bar{D} \hat{D},
\end{equation}
into (\ref{eqn:advdiff_final}) we obtain
\begin{equation}
\frac{c_0}{\delta t} \frac{\partial \hat{c}}{\partial  \hat{t}}  + \frac{U c_0}{L} \hat{u}_F \frac{\partial \hat{c}}{\partial \hat{x} }+ \frac{W c_0}{2R} \frac{\partial (\hat{v}_p \hat{c})}{\partial \hat{y}}  =
\frac{ \bar{D} c_0  }{L^2} \left(\hat{D} \frac{\partial^2 \hat{c}}{\partial \hat{x}^2} \right)+
\frac{\bar{D} c_0 }{4 R^2} \frac{\partial }{\partial \hat{y}}\left( \hat{D} \frac{\partial \hat{c}}{\partial \hat{y}}\right)\,,
\end{equation}
where $U=\max(u_F)$, $W=\max(v_p)$ and $\bar{D}=\max(D)$, vary slightly depending on the model of fluid chosen. As a reference, for the Newtonian fluid with $F_0 = 0.5 \times 10^{-14}$\,N, we have $U=7.5\cdot10^{-4}$\,m/s, $W=5.13\cdot10^{-6}$\,m/s and $D=1.37\cdot10^{-10}$\,m$^{2}$/s. The length scales are $L=500$\,$\mu$m and $2R = 40$\,$\mu$m. 

Rearranging the terms and choosing $\delta t = L / U$ to balance the time derivative with the advection term we obtain
\begin{equation}\label{eqn:advdiff_epsilon}
\frac{\partial \hat{c}}{\partial  \hat{t}}  + \hat{u}_F \frac{\partial \hat{c}}{\partial \hat{x} }+ \frac{\delta}{\varepsilon}\frac{\partial (\hat{v}_p \hat{c})}{\partial \hat{y}} = \frac{1}{ \varepsilon \, \mathrm{Pe}} \left[
\varepsilon^2 \left(\hat{D} \frac{\partial^2 \hat{c}}{\partial \hat{x}^2} \right)+
\frac{\partial }{\partial \hat{y}}\left( \hat{D} \frac{\partial \hat{c}}{\partial \hat{y}}\right)\right],
\end{equation}
where $\delta = W/U$ and $\mathrm{Pe} =  2R U/D $ is the P\'eclet number. $\mathrm{Pe}$ depends on the model chosen. The time scale $\delta t$ represents the average time taken by a particle to pass through the vessel in the absence of a magnetic field (e.g., in the Newtonian case $\delta t=6.67$\,s). According to our choice of scales and the values in Table \ref{Table:PhysicalParameters_conc}, $\mathcal{O}( (\varepsilon \,\mathrm{Pe})^{-1}) \approx 10^{-1} $ depending on the fluid chosen, while $\varepsilon = \mathcal{O}(10^{-2})$. Hence, both terms on the right-hand side of \eqref{eqn:advdiff_epsilon} are  small. Therefore, the advective terms are dominant and when analysing them, it is important to understand the order of magnitude of the fraction $\delta/\varepsilon$. In particular we can distinguish three regions in the domain (symmetric with respect to the center of the vessel): a central region where $\mathcal{O}(\delta) < \mathcal{O}(\varepsilon)$, which is the broadest one where the drag force dominates over the magnetic force; a second region where $\mathcal{O}(\delta) \approx \mathcal{O}(\varepsilon)$ where both advective terms  balance; finally, very near to the wall of the vessel, we can find a narrow boundary layer where $\mathcal{O}(\delta) > \mathcal{O}(\varepsilon)$ (Stokes drag is very small) and vertical motion due to the magnetic field dominates.

The boundary conditions at the vessel wall are 
\begin{align}\label{wall}
\hat{v}_p \hat{c}-\frac{1}{\mathrm{Pe}\,\delta}\hat{D}\frac{\partial \hat{c}}{\partial \hat{y}}=\pm h \hat{c}\qquad \text{on}\qquad \hat{y} = \pm \frac{1}{2}\,,
\end{align}
where $h=\kappa/W$. At the vessel inlet 
\begin{align}\label{inlet}
\hat{u}_F \hat{c}_{\text{in}}= \hat{u}_F \hat{c}-\frac{\varepsilon} {\mathrm{Pe}}\hat{D}\frac{\partial \hat{c}}{\partial \hat{x}} \qquad \text{on}\qquad \hat{x} = 0\,. 
\end{align}
Finally, at the outlet we have 
\begin{align}\label{final}
\hat{u}_F \hat{c}-\frac{\varepsilon} {\mathrm{Pe}}\hat{D}\frac{\partial \hat{c}}{\partial \hat{x}} =0 \qquad \text{on}\qquad \hat{x} = 1\,, 
\end{align}
which does not affect the results since the simulations are always stopped when the particles are still far from reaching the vessel outlet. This ensures that nanoparticles only leave the vessel through absorption at the boundaries $y=\pm 1/2$ during our simulations.

\section{Appendix C: Finite difference scheme for the diffusion equation}
\label{subs:numericalscheme}

We drop the hats in \eqref{eqn:advdiff_epsilon}-\eqref{final} and define
\begin{equation}
c_{i,j}^n \coloneqq c(x_i,y_j, t^n), \qquad u_{F_j} \coloneqq u_F(y_j)  , \qquad v_{p_{j}} \coloneqq v_p(y_j), \qquad D_{\mathrm{T}_{j}}\coloneqq  D(y_j).
\end{equation}
The choice of the direction of the upwind step is made considering that the solution of the velocity of the fluid is always non negative and the direction of the vertical velocity is always negative (since we have positioned the magnet below the vein). Then, equation (\ref{eqn:advdiff_epsilon}) can be approximated as
\begin{align}\label{eqn:completescheme}
\begin{split}
& \frac{c_{i,j}^{n+1} - c_{i,j}^n}{\Delta t} + \frac{\delta}{\varepsilon}\, u_{F_j} \left( \frac{c_{i,j}^{n} - c_{i-1,j}^{n}}{\Delta x} \right) +
\left(\frac{v_{p_{j+1}}c_{i,j+1}^{n} - v_{p_{j}}c_{i,j}^{n}}{\Delta y} \right) = \\
& \frac{\varepsilon}{ \mathrm{Pe}}  D_{\mathrm{T}_{j}} \left( \frac{c_{i+1,j}^{n} -2 c_{i,j}^{n} + c_{i-1,j}^{n}}{\Delta x^2}\right) + \frac{1}{ \varepsilon \, \mathrm{Pe}}  D_{\mathrm{T}_{j+\frac{1}{2}}} \left(  \frac{c_{i,j+1}^{n} - c_{i,j}^{n} }{\Delta y^2} \right)
\\ 
&  - \frac{1}{ \varepsilon \, \mathrm{Pe}}  D_{\mathrm{T}_{j-\frac{1}{2}}} \left( \frac{c_{i,j}^{n} - c_{i,j-1}^{n}}{\Delta y^2} \right),
\end{split}
\end{align}
where $D_{\mathrm{T}_{j \pm \frac{1}{2}}}$ are evaluated by the arithmetic mean. 

We also need to specify the boundary conditions at $x = 0,\,1$ and $y=\pm 1/2$. On the wall of the vessel, that is the upper and the lower side of the rectangle, we approximated conditions (\ref{wall}), through the three-point backward difference formula
\begin{equation}\label{eqn:bcup}
v_{p_{n_y}} \,  c_{i,n_y}^{n+1}  - \frac{1}{\mathrm{Pe}\,\delta} D_{\mathrm{T}_{n_y}} \left(\frac{3 c_{i,n_y}^{n+1} - 4 c_{i,n_y-1}^{n+1} + c_{i,n_y-2}^{n+1}}{2 \Delta y} \right) = h\, c_{i,n_y}^{n+1}\,,
\end{equation}
and the three-point forward difference formula
\begin{equation}\label{eqn:bcbot}
v_{p_1} \, c_{i,1}^{n+1} - \frac{1}{\mathrm{Pe}\,\delta} D_{\mathrm{T}_{1}}  \left(\frac{ - c_{i,3}^{n+1} + 4 c_{i,2}^{n+1} -3 c_{i,1}^{n+1} }{2 \Delta y} \right) = -h\, c_{i,1}^{n+1}\,,
\end{equation}
for $i= 1,\dots,n_x$. Similarly, the boundary conditions at the vessel inlet and outlet are approximated again through the three-point difference formulas
\begin{equation}\label{eqn:bcleft}
u_{F_j} c_{\text{in},j}^{n+1} = u_{F_j} c_{1,j}^{n+1} - \frac{\varepsilon} {\mathrm{Pe}} D_{\mathrm{T}_{j}} 
\left(\frac{ - c_{3, j}^{n+1} + 4 c_{2, j}^{n+1} -3 c_{1,j}^{n+1} }{2 \Delta x} \right) \,,
\end{equation}
\begin{equation}\label{eqn:bcright}
u_{F_j} c_{n_x,j}^{n+1} - \frac{\varepsilon} {\mathrm{Pe}} D_{\mathrm{T}_{j}} \left(\frac{ - c_{n_x-2, j}^{n+1} + 4 c_{n_x-1, j}^{n+1} -3 c_{n_x,j}^{n+1} }{2 \Delta x} \right)  = 0\,,
\end{equation}
for $j = 1,\dots,n_y$.


\bibliographystyle{unsrt}
\bibliography{nanodrug_bib}

\end{document}